\documentclass{appolb}
\usepackage{epsfig}

\begin{document}
\title{LARGE-$\theta_{13}$ PERTURBATION THEORY OF NEUTRINO OSCILLATION\thanks{An expanded written version of the talk presented at 
XXXIII International Conference of Theoretical Physics 
``MATTER TO THE DEEPEST'': Recent Developments in Physics 
of Fundamental Interactions (Ustron'09). 
}%
}
\author{Hisakazu MINAKATA
\address{Department of Physics, Tokyo Metropolitan University, Minami-Osawa, 
Hachioji, Tokyo 192-0397, Japan}
}
\maketitle

\begin{abstract} 
Keeping in mind the possibility of large $\theta_{13}$, which will be soon explored 
by reactor and accelerator experiments, I formulate a perturbation theory of neutrino 
oscillation under the ansatz 
$s_{13}^2 \simeq \Delta m^2_{21} / \Delta m^2_{31}  \simeq 0.03$, 
which is comparable to the Chooz limit. 
Under the framework, I derive the perturbative formula of the $\nu_{e}$ appearance 
probability valid to order $\epsilon^2$ in which effects of arbitrary matter density 
profile is taken into account.
I use the formula to analyze problem of possible obstruction to detecting 
lepton CP violation by effects of asymmetry in matter density profile. 
Though the asymmetry could be large for neutrino trajectories which traverse both 
continental and sea crust, its effect on obscuring CP violation 
measurement is found to be quite small. 

\end{abstract}

\section{Introduction}

My presentation in Ustron'09 was done under the title 
``Long-Baseline (LBL) Neutrinos; Looking Forward to the Future''. 
It included a review of the ideas for exploration of the unknowns in the 1-3 sector 
of the MNS matrix \cite{MNS}, CP violation due to the lepton analogue of the Kobayashi-Maskawa phase \cite{KM} $\delta$ and the neutrino mass hierarchy. 
If I restrict myself into perspective in North-East Asia, they include an upgrade 
of J-PARC beam with megaton-scale Hyper-Kamiokande (T2K II) 
as described in \cite{T2K}, 
a 100 kt scale liquid Ar detector in Okinoshima \cite{okinoshima}, and the 
Tokai-to-Kamioka-Korea (for short T2KK) setting \cite{T2KK1st,T2KK2nd}. 
The overview of the latter is given in \cite{T2KK-review}. 
In particular, I emphasized the Kamioka-Korea identical two-detector setting 
as a robust way of measuring the CP violating phase and determining the 
mass hierarchy. 
It combines the idea of low energy superbeam as the cleanest way for detecting 
lepton CP violation \cite{superbeam} and the powerfulness of the two-detector 
method \cite{MNplb97}. 
But, since the contents of this part are described in the previous reports  
\cite{HQL06,nu2008-mina}
I confine myself into the last part of my presentation in Ustron'09, 
the large-$\theta_{13}$ perturbation theory in this written version.

\section{Motivation and Use}

The motivation for formulating the large-$\theta_{13}$ perturbation theory is very simple; 
$\theta_{13}$ can be as large as the Chooz limit \cite{CHOOZ}. 
If it is the case $s_{13} \simeq 0.17$. 
I emphasize that this possibility is to be tested very soon by the accelerator 
\cite{T2K,NOVA} and the reactor $\theta_{13}$ experiments \cite{reactor-exp} 
some of which will start in 2009. 
Then, the $\epsilon$ perturbation theory 
(in a terminology defined in \cite{NSI-perturbation}) 
formulated under the ansatz 
$s_{13} \simeq \epsilon \equiv \Delta m^2_{21} / \Delta m^2_{31}  \simeq 0.03$, 
which provides the simplest way of deriving the widely used Cervera {\it et al.} 
formula \cite{golden} of the oscillation probability, 
would not serve as the best approximation.

Then, what is the use of the large-$\theta_{13}$ perturbation theory? 
I confine in this paper the robustness issue in uncovering lepton CP violation. 
As is well recognized the matter effect produces a fake CP violation. 
The presence of the matter effect is inevitable for settings which also 
have sensitivity to the mass hierarchy. Early references of this topics include \cite{AKS,MNprd98}. 
The problem has been discussed in a number of authors which produced 
too many references to quote here. 
However, there exists a point which does not appear to be given full attention in the literature, the problem of possible asymmetry in 
matter density profile in the earth. 
See, however, \cite{akhmedov99} for discussion of this problem. 
It is known that asymmetric baseline produces CP violating $\sin \delta$ terms in 
$P(\nu_{\mu} \rightarrow \nu_{\mu})$ \cite{solarCP}, and $\cos \delta$ terms 
in T violating observable $\Delta P_{T}$ \cite{yokomakura} which invalidates 
the neat property of T violation,  
$\Delta P_{T}=0$ for vanishing $\delta$ \cite{petcov88}.
Apparently, the effects of asymmetric matter density profile is most 
prominent for large $\theta_{13}$.

A crucial question would be: 
Are there any situation in which sizable asymmetry in matter density profile 
shows up? The answer is indeed {\em yes}.
Suppose a baseline of order $\sim 1000$ km and a neutrino beam launched 
at a place on a continent is received by a detector which is placed in an island. 
If the travel distances of the beam in the continent and in the sea are comparable, 
one can expect asymmetry in the density profile because the matter density 
in the earth crust is believed to be smaller under the sea. 
Contrary to the mantle density, the crust density is not severely restricted by 
the earth mass, and it would not be easy to measure it directly either. 
Therefore, it is important to investigate to what extent CP violation discovery 
might be obscured by asymmetry in matter density profile.

\section{Large-$\theta_{13}$ perturbation theory }
\label{L-perturbation}

The neutrino evolution equation can be written in flavor basis as 
$ i \frac{ d }{ dx } \nu_{\alpha} = H_{ \alpha \beta } \nu_{ \beta }~(\alpha, \beta = e, \mu, \tau)$. 
In the standard three-flavor neutrino scheme, Hamiltonian is given by 
\begin{eqnarray}
H= 
\frac{1}{2E} 
\left\{ 
U \left[
\begin{array}{ccc}
0 & 0 & 0 \\
0 & \Delta m^2_{21} & 0 \\
0 & 0 & \Delta m^2_{31} 
\end{array}
\right] U^{\dagger}
+ 
a(x) \left[
\begin{array}{ccc}
1 & 0 & 0 \\
0 & 0 & 0 \\
0 & 0 & 0
\end{array}
\right] 
\right\} 
\label{hamiltonian}
\end{eqnarray}
where $\Delta m^2_{ji} \equiv m^2_{j} - m^2_{i}$, and 
$a(x) \equiv 2\sqrt{2} G_F N_e(x) E$ is the coefficient which is 
related to the index of refraction of neutrinos in medium of electron 
number density $N_e(x)$, where $G_F$ is the Fermi constant and 
$E$ is the neutrino energy. 
$U = U_{23} U_{13} U_{12} $ is the MNS matrix \cite{MNS} in the lepton sector.


It is straightforward to formulate the perturbative framework of neutrino oscillations. 
We refer \cite{NSI-perturbation} for notations. 
We use the tilde-basis 
$ \tilde{\nu} = U_{23}^{\dagger} \nu $ with the tilde-Hamiltonian 
$ \tilde{H} = U_{23}^{\dagger}  H  U_{23} $, which is decomposed 
as $ \tilde{H} = \tilde{H}_{0} + \tilde{H}_{1} $. 
Then, the $S$ matrix can be written as 
\begin{eqnarray} 
S(L) = U_{23} e^{- i \tilde{H}_{0} x}  \Omega(x)  U_{23}^{\dagger}. 
\label{Smatrix}
\end{eqnarray}
$\Omega(x)$ can be expanded with use of 
$ H_{1} \equiv e^{i \tilde{H}_{0} x} \tilde{H}_{1} e^{-i \tilde{H}_{0} x} $ as 
\begin{eqnarray} 
\Omega(x) = 1 + 
(-i) \int^{x}_{0} dx' H_{1} (x') + 
(-i)^2 \int^{x}_{0} dx' H_{1} (x') \int^{x'}_{0} dx'' H_{1} (x'') + 
\mathcal{O} ( \epsilon^3 ) 
\label{Omega-exp}
\end{eqnarray}
where the ``space-ordered'' form in (\ref{Omega-exp}) is essential 
because of the highly nontrivial spatial dependence in $H_{1}$. 

We use the method of Fourier decomposition to incorporate the effect of 
matter density variation in the earth \cite{ota-sato01} with the dimensionless 
variable $r_{A} \equiv \frac{ a }{ \Delta m^2_{31} } $. 
It can be expanded into a Fourier series as 
\begin{eqnarray} 
r_{A} (x)  = 
r^{A}_{0} + 
\sum^{ \infty }_{n=1} \biggl[ 
r^{A}_{n} e^{ - i p_{n} x } + (r^{A}_{n})^* e^{  i p_{n} x } 
\biggr]
\label{rA-fourier}
\end{eqnarray}
where $p_{n} \equiv \frac{ 2\pi }{ L } n$. 
It should be noticed that if $r_{A} (L - x) = r_{A} (x)$, namely if the baseline 
is symmetric, $r^{A}_{n} = (r^{A}_{n})^* $. 
Therefore, the imaginary part of $r^{A}_{n}$ represents the effect of 
asymmetric matter density profile.


We assume large $\theta_{13}$ comparable to the Chooz limit, 
$s_{13} \simeq \sqrt{\epsilon}$, where 
$\epsilon \equiv \Delta m^2_{21} / \Delta m^2_{31} \simeq 0.03$. 
Other small parameters are present depending upon experimental 
setting, in combination with matter density, neutrino energy, and baseline. 
As a model superbeam experiment we consider a baseline with distance 
$L=1000$ km and neutrino energy $E=2$ GeV. 
Noticing that 
$\frac{a} { \Delta m^2_{31} } = 0.084
\left(\frac{ \Delta m^2_{31} }{ 2.5 \times 10^{-3}\mbox{eV}^2}\right)^{-1}
\left(\frac{E}{1\mbox{GeV}}\right)
\left(\frac{\rho}{2.8 \mbox{g/cm}^3}\right)$,
%
%
%
the setting leads to $ a / \Delta m^2_{31} = 0.17  \simeq \sqrt { \epsilon}$.
(For T2K II setting, $ a / \Delta m^2_{31} \simeq \epsilon$ may be more appropriate.) 
Then, we formulate the large-$\theta_{13}$ perturbation theory by taking 
the following expansion parameters 
$\epsilon = \Delta m^2_{21} / \Delta m^2_{31} $ and 
$ s_{13} \simeq a / \Delta m^2_{31} \simeq \sqrt { \epsilon}$. 
The unperturbed part of the tilde-basis Hamiltonian is given by 
$\tilde{H}_{0} = \Delta \mbox{diag} (0, 0, 1)$, while the perturbed part is 
\begin{eqnarray} 
\tilde{H}_{1} &=& 
\Delta \left\{ 
\left[
\begin{array}{ccc}
r_{A} (x) & 0 & s_{13} e^{ -i \delta} \\
0 & 0 & 0 \\
s_{13} e^{ i \delta} & 0 & 0 
\end{array}
\right] 
+
\left[
\begin{array}{ccc}
\epsilon s^2_{12} + s^2_{13} & \epsilon  c_{12} s_{12}  & 0 \\
\epsilon  c_{12} s_{12}  & \epsilon  c^2_{12}  & 0 \\
0 & 0 & - s^2_{13}  
\end{array}
\right] \right\} 
\nonumber \\
&-&  
\Delta \epsilon 
\left[
\begin{array}{ccc}
0  & 0 & s^2_{12} s_{13} e^{ -i \delta}  \\
0 & 0 & c_{12} s_{12} s_{13} e^{ -i \delta}   \\
s^2_{12} s_{13} e^{ i \delta}  & c_{12} s_{12} s_{13} e^{ i \delta}  & 0  
\end{array}
\right] 
\nonumber \\ 
&-&
\Delta \epsilon  \left[
\begin{array}{ccc}
 s^2_{12} s^2_{13} & \frac{1}{2} c_{12} s_{12} s^2_{13} & 0 \\
 \frac{1}{2} c_{12} s_{12} s^2_{13} & c^2_{12}  & 0 \\
0 & 0 & - s^2_{12} s^2_{13}
\end{array}
\right], 
\label{H1}
\end{eqnarray}
where 
$\Delta = \Delta m^2_{31} / 2E $. 
The first, second, third, and the fourth terms in (\ref{H1}) are of order 
$\epsilon^{\frac{1}{2}}$, 
$\epsilon^{1}$, 
$\epsilon^{\frac{3}{2}}$, and 
$\epsilon^{2}$, respectively.

The S matrix can be computed with use of (\ref{Smatrix}). 
Then, the appearance oscillation probability 
$P(\nu_{\mu} \rightarrow \nu_{e}) = \vert S_{e \mu} \vert^2 $ 
is given to second order in $\epsilon$ 
with a simplified notation $\Delta_{31} \equiv \Delta m^2_{31} L / 4E $ by 
\begin{eqnarray} 
&& \hspace{-2mm}  P(\nu_{\mu} \rightarrow \nu_{e}) = 
4 s^2_{23} s^2_{13} \sin^2 \Delta_{31} 
\left[ 
\left\{ 1 + r^{A}_{0} + 2 \sum^{ \infty }_{ n=1 } \frac{ \Delta_{31}^2 }{ \Delta_{31}^2 - (n\pi)^2 } \mbox{Re}  \left( r^{A}_{n} \right) \right\}^2 
%
\right.
\nonumber \\
&&\hspace*{3mm} {} +
\left. 
s^2_{13} ( 3 - 4 \Delta_{31} \cos \Delta_{31} ) 
%
\right.
\nonumber \\
&&\hspace*{3mm} {} +
\left. 
4 \Delta_{31}^2  
 \sum^{ \infty }_{ n=1 } \frac{ n\pi }{ \Delta_{31}^2 - (n\pi)^2 } \mbox{Im}  \left( r^{A}_{n} \right) 
\left\{ r^{A}_{0} 
+ \sum^{ \infty }_{ n=1 }  \frac{ n\pi }{ \Delta_{31}^2 - (n\pi)^2 } \mbox{Im}  \left( r^{A}_{n} \right) 
\right\} \right]  
\nonumber \\
&-& 
4 s^2_{23} s^2_{13} \Delta_{31} \sin 2 \Delta_{31} 
\left [ 
\epsilon s^2_{12} + r^{A}_{0} 
\left\{ 1 + r^{A}_{0} 
+ 2 \sum^{ \infty }_{ n=1 } \frac{ \Delta_{31}^2 }{ \Delta_{31}^2 - (n\pi)^2 } \mbox{Re}  \left( r^{A}_{n} \right) \right\}
\right] 
\nonumber \\
&+& 
4 \Delta_{31}^2  
\left[ 
\epsilon^2 c^2_{12} s^2_{12} c^2_{23} + s^2_{23} s^2_{13} (r^{A}_{0})^2 \right] 
- 8  J_r \epsilon r^{A}_{0} \Delta_{31}^2 \cos \delta 
\nonumber \\
%
&+& 
8 J_r \epsilon \Delta_{31} 
\sin \Delta_{31} 
\cos \left( \delta + \Delta_{31} \right) 
\left[ 1 + r^{A}_{0} + 
2 \sum^{ \infty }_{ n=1 } \frac{ \Delta_{31}^2 }{ \Delta_{31}^2 - (n\pi)^2 } \mbox{Re}  \left( r^{A}_{n} \right) 
\right]
\nonumber \\
&+& 
8 J_r \epsilon \Delta_{31}^2
\sin \Delta_{31} 
\sin \left( \delta +  \Delta_{31} \right) 
\left [ r^{A}_{0} 
- 2  \sum^{ \infty }_{ n=1 } \frac{ \Delta_{31}^2 }{  n\pi \left[ \Delta_{31}^2 - (n\pi)^2 \right] } \mbox{Im}  \left( r^{A}_{n} \right) 
\right]. 
\label{Pemu}
\end{eqnarray}

\section{T, CP and CPT violation observable }

It is interesting to compute the expression of T, CP and CPT violation 
observable to know the difference in their dependence on $\delta$ and the 
matter density profile. 
The T, CP, and CPT conjugate probabilities can be obtained by the 
appropriate replacements:  \\
$P(\nu_{e} \rightarrow \nu_{\mu})  = 
P \left( \nu_{\mu} \rightarrow \nu_{e};  - \delta, r^{A}_{0}, \mbox{Re} ( r^{A}_{n} ), - \mbox{Im}  ( r^{A}_{n} ) \right)$, \\
$P(\bar{\nu}_{\mu} \rightarrow \bar{\nu}_{e})  = 
P \left( \nu_{\mu} \rightarrow \nu_{e};  - \delta, - r^{A}_{0}, - \mbox{Re} ( r^{A}_{n} ), - \mbox{Im} ( r^{A}_{n} ) \right)$, \\
$P(\bar{\nu}_{e} \rightarrow \bar{\nu}_{\mu})  = 
P \left( \nu_{\mu} \rightarrow \nu_{e};  \delta, - r^{A}_{0}, - \mbox{Re} ( r^{A}_{n} ), \mbox{Im} ( r^{A}_{n} ) \right)$. \\
With (\ref{Pemu}) it is  it is easy to 
calculate them and the results read: 
%
%
\begin{eqnarray} 
&& \hspace{-14mm}
\mbox{T violation}:   \Delta P_{T} \equiv 
P(\nu_{\mu} \rightarrow \nu_{e}) - P(\nu_{e} \rightarrow \nu_{\mu}) 
\nonumber \\
&=& 
- 16 J_r  \sin \delta  \epsilon \Delta_{31} 
\sin^2 \Delta_{31} 
\nonumber \\
&-& 
16 J_r \sin \delta \epsilon \Delta_{31}  
\left(
\sin^2 \Delta_{31} 
- \frac{1}{2} \Delta_{31}
\sin 2 \Delta_{31} 
\right) 
r^{A}_{0} 
\nonumber \\
&-& 
32 J_r \sin \delta \epsilon \Delta_{31} 
\sin^2 \Delta_{31}  
\sum^{ \infty }_{ n=1 } \frac{ \Delta_{31}^2 }{ \Delta_{31}^2 - (n\pi)^2 } \mbox{Re}  \left( r^{A}_{n} \right) 
\nonumber \\
&-& 
32 J_r  \cos \delta \epsilon \Delta_{31}^2
\sin^2 \Delta_{31}  
\sum^{ \infty }_{ n=1 } \frac{ \Delta_{31}^2 }{  n\pi \left[ \Delta_{31}^2 - (n\pi)^2 \right] } \mbox{Im}  \left( r^{A}_{n} \right) 
\nonumber \\
&+& 
32 s^2_{23} s^2_{13} \Delta_{31}^2  \sin^2 \Delta_{31} r^{A}_{0} 
 \sum^{ \infty }_{ n=1 } \frac{ n\pi }{ \Delta_{31}^2 - (n\pi)^2 } \mbox{Im}  \left( r^{A}_{n} \right). 
\label{Pmue-Tviolation}
\end{eqnarray}
%
%
\begin{eqnarray} 
&& \hspace{-14mm}
\mbox{CP violation}:  \Delta P_{CP} \equiv 
P(\nu_{\mu} \rightarrow \nu_{e}) - P(\bar{\nu}_{\mu} \rightarrow \bar{\nu}_{e}) 
\nonumber \\
&=& -  
16 J_r \sin \delta \epsilon \Delta_{31} \sin^2 \Delta_{31} 
\nonumber \\
&+& 
16 s^2_{23} s^2_{13} 
\left(
\sin^2 \Delta_{31} 
- \frac{1}{2} \Delta_{31} \sin 2 \Delta_{31} 
\right) r^{A}_{0} 
\nonumber \\
&+&
32 s^2_{23} s^2_{13} \sin^2 \Delta_{31} 
\sum^{ \infty }_{ n=1 } \frac{ \Delta_{31}^2 }{ \Delta_{31}^2 - (n\pi)^2 } \mbox{Re}  \left( r^{A}_{n} \right) 
\nonumber \\
&+& 
16 J_r \cos \delta \epsilon \Delta_{31} 
\left(
\Delta_{31} \sin^2 \Delta_{31} 
+ \frac{1}{2} \sin 2 \Delta_{31} 
-  \Delta_{31} 
\right)  r^{A}_{0} 
\nonumber \\
&+& 
16 J_r \cos \delta \epsilon \Delta_{31} 
\sin 2 \Delta_{31}  
\sum^{ \infty }_{ n=1 } \frac{ \Delta_{31}^2 }{ \Delta_{31}^2 - (n\pi)^2 } \mbox{Re}  \left( r^{A}_{n} \right) 
\nonumber \\
&-& 
32 J_r \cos \delta \epsilon \Delta_{31}^2 \sin^2 \Delta_{31}  
\sum^{ \infty }_{ n=1 } \frac{ \Delta_{31}^2 }{  n\pi \left[ \Delta_{31}^2 - (n\pi)^2 \right] } \mbox{Im}  \left( r^{A}_{n} \right). 
\label{Pmue-CPviolation}
\end{eqnarray}
%
%
\begin{eqnarray} 
&& \hspace{-7mm}
\mbox{CPT violation}:  
\Delta P_{CPT} \equiv 
P(\nu_{\mu} \rightarrow \nu_{e}) - P(\bar{\nu}_{e} \rightarrow \bar{\nu}_{\mu}) 
\nonumber \\
&=& 
16 s^2_{23} s^2_{13} 
\left( \sin^2 \Delta_{31} 
- \frac{1}{2} \Delta_{31} \sin 2 \Delta_{31} 
\right) r^{A}_{0} 
\nonumber \\
&+& 
32 s^2_{23} s^2_{13} \sin^2 \Delta_{31} 
\sum^{ \infty }_{ n=1 } \frac{ \Delta_{31}^2 }{ \Delta_{31}^2 - (n\pi)^2 } \mbox{Re}  \left( r^{A}_{n} \right)  
\nonumber \\
&+& 
16 J_r \epsilon \Delta_{31}  
\biggl[
\Delta_{31} \sin \Delta_{31} \sin \left( \delta +  \Delta_{31} \right) 
+ \sin \Delta_{31} \cos \left( \delta +  \Delta_{31} \right) 
- \Delta_{31} \cos \delta 
\biggr] 
r^{A}_{0} 
\nonumber \\
&+& 
32 J_r \epsilon \Delta_{31} 
\sin \Delta_{31} 
\cos \left( \delta + \Delta_{31} \right) 
\sum^{ \infty }_{ n=1 } \frac{ \Delta_{31}^2 }{ \Delta_{31}^2 - (n\pi)^2 } \mbox{Re}  \left( r^{A}_{n} \right) 
\nonumber \\
&+& 
32 s^2_{23} s^2_{13} \Delta_{31}^2  \sin^2 \Delta_{31} 
r^{A}_{0} 
 \sum^{ \infty }_{ n=1 } \frac{ n\pi }{ \Delta_{31}^2 - (n\pi)^2 } \mbox{Im}  \left( r^{A}_{n} \right). 
\label{Pmue-CPTviolation}
\end{eqnarray}
In (\ref{Pmue-Tviolation}), (\ref{Pmue-CPviolation}), and (\ref{Pmue-CPTviolation}), 
first a few terms are of order $\epsilon^{3/2}$, and the rests are of order 
$\epsilon^{2}$. The effects of asymmetric profile are always contained in the latter. 
As we noted earlier the charming property of T violation observable holds 
without asymmetry in the matter profile, $\mbox{Im} (r^{A}_{n})=0$. 
That is, if $\delta$ vanishes then $\Delta P_{T} = 0$; 
The matter effect cannot produce a fake T violation. 
The asymmetric baseline destroys the neat property.

We examine the effect of asymmetry in the matter density profile by taking 
an explicit model of the profile:
\begin{eqnarray} 
\rho &=& \rho_{0} + \delta \rho 
\hspace{10mm} (0 \leq x \leq L/2 ), 
\nonumber \\
\rho &=& \rho_{0} - \delta \rho
\hspace{10mm} (L/2 \leq x \leq L), 
\label{profile}
\end{eqnarray}
which leads to 
$\mbox{Im} (r^{A}_{n}) / r^{A}_{0} = (2/\pi n) \delta \rho / \rho_{0}$ (odd n), and 
$\mbox{Re} (r^{A}_{n}) =0$. 
$r^{A}_{0} = a_{0} / \Delta m^2_{31}$ where $a_{0}$ is obtained by using the 
matter density $\rho_{0}$ in the definition of $a$. 
For concreteness we take $ \rho_{0} = 2 g/\mbox{cm}^3$ and 
$ \delta \rho = 0.8 g/\mbox{cm}^3$. 
We use the toy model of matter density profile to compute T, CP, and CPT violating observable $\Delta P_{T}$, $\Delta P_{CP}$, 
and $\Delta P_{CPT}$ defined respectively in (\ref{Pmue-Tviolation}), 
(\ref{Pmue-CPviolation}), and (\ref{Pmue-CPTviolation}). 
We do this under the approximation of keeping only the lowest mode 
$\mbox{Im} (r^{A}_{1})$ ($n=1$), which appears to give a good approximation. 
I take $\sin^2 2\theta_{13}=0.1$ and $\delta=3\pi/4$ in this calculation, 
but it appears that qualitative features are rather insensitive to $\delta$.

\begin{figure}[htb]
\begin{center}
\vspace{-0.4cm}
\epsfig{file=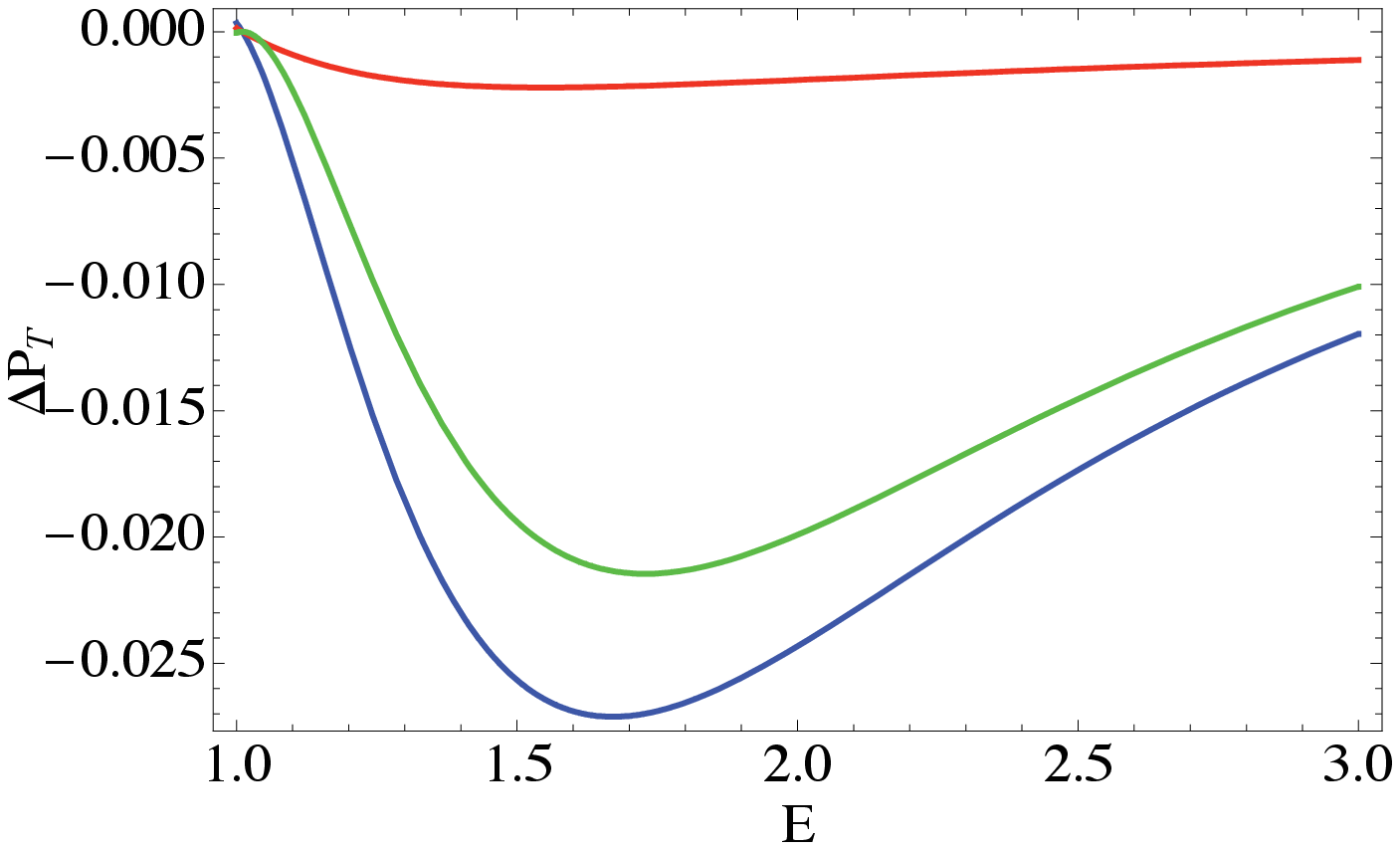,width=3.1 in}
\epsfig{file=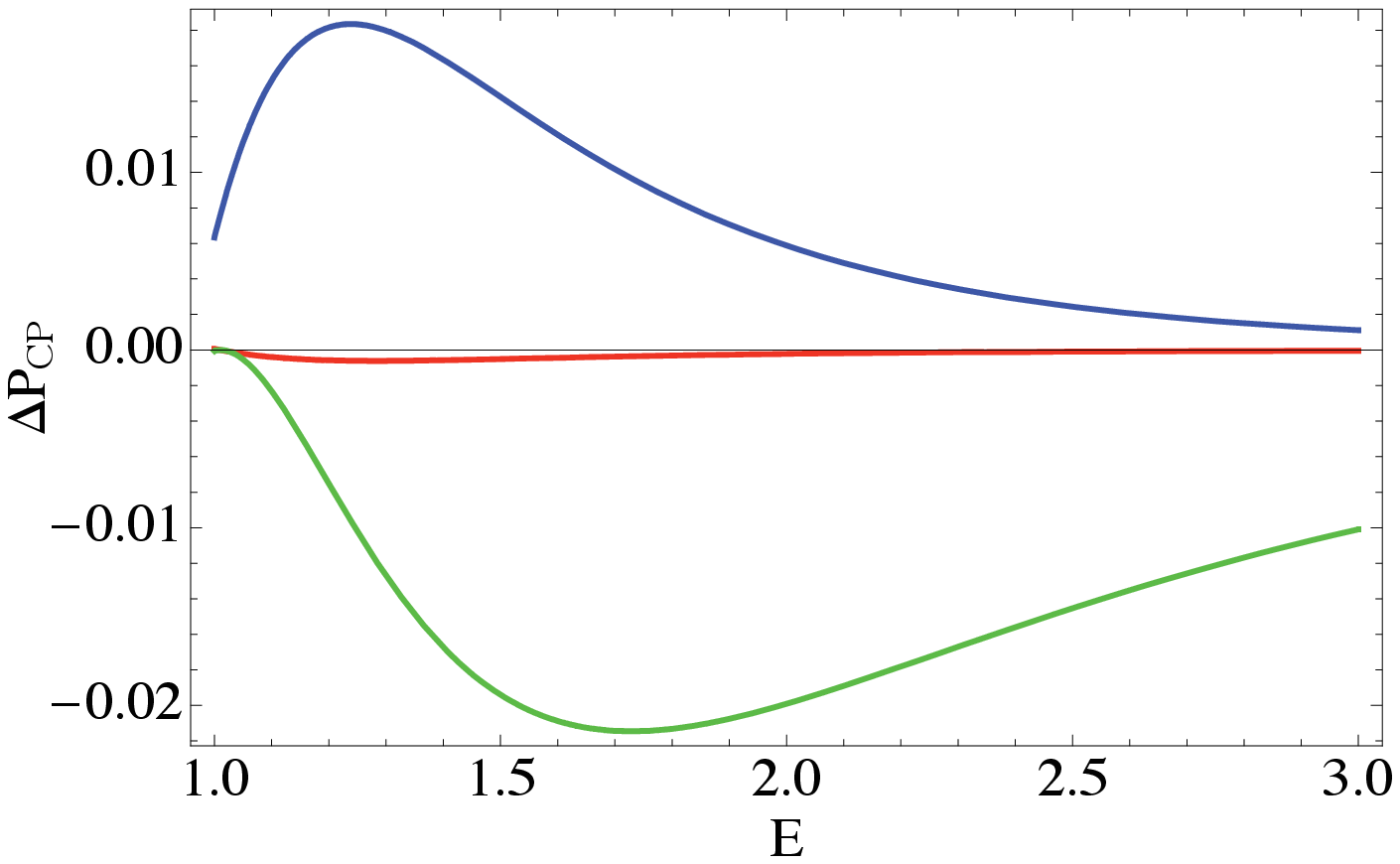,width=3.1 in}
\epsfig{file=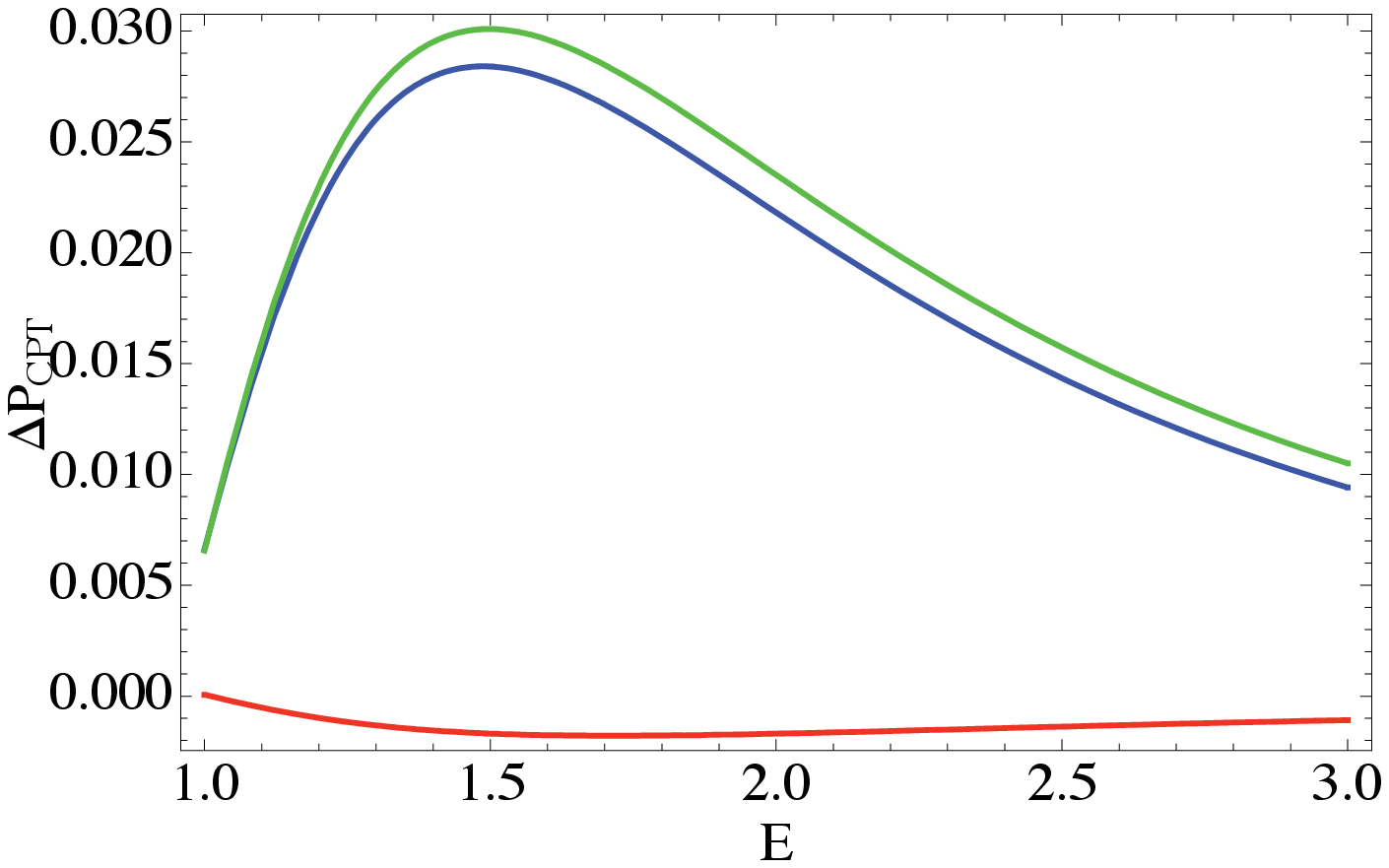,width=3.1 in}
\caption{T, CP, and CPT violating observable $\Delta P_{T}$, $\Delta P_{CP}$, and $\Delta P_{CPT}$ defined respectively in (\ref{Pmue-Tviolation}), 
(\ref{Pmue-CPviolation}), and (\ref{Pmue-CPTviolation}) are plotted as a function of neutrino energy (in GeV) in the upper, middle, and the lower panels, respectively.  
In each panel the blue and the red lines indicate, respectively, 
the value of $\Delta P$ and the contribution to $\Delta P$ from 
$\mbox{Im} (r^{A}_{1})$ terms. 
The green lines in the upper and the middle panels indicate the contribution 
of vacuum effect only. The green line in the bottom panel represents the 
contribution from average matter density.  
}
\label{DeltaP-T-CP-CPT}
\vspace{-0.2cm}
\end{center}
\end{figure}

The results of the calculation are presented in Fig.~\ref{DeltaP-T-CP-CPT}. 
One can see immediately that, in spite of a rather large asymmetry in density 
profile (\ref{profile}) the effect of the asymmetric baseline (shown by the red lines) 
is small. In fact, it is negligibly small in CP violation observable. 
It appears that this feature prevails for other choices of CP violating phase $\delta$.
I have also checked that the effect of the asymmetry becomes even smaller for 
shorter baseline and/or smaller $\theta_{13}$. 
I believe that this settles the issue of possible contamination effects by 
the asymmetric baseline to detection of genuine CP violation effect; 
A good news for medium baseline ($\sim 1000$ km) experiments.

Some remarks are in order:

\begin{itemize}

\item 
The size of the matter effect (difference between blue and green lines) gives a 
dominant effect in $\Delta P_{CP}$, overturning the negative sign of the vacuum contribution in this particular case. But, it plays only a minor role in $\Delta P_{T}$. 
CPT-violation observable would be most powerful to resolve the mass hierarchy 
\cite{MNP3} because it gives the largest $\Delta P$.

\item

In CP (also T) violation observable, the energy dependence of the average matter 
density term (blue minus (green + red)) are rather similar to the vacuum term. 
Then, there could be severe confusion between CP violation caused 
by phase and uncertainty in the average density of matter. 
In particular, it gives rise to a serious confusion at $\delta=0$ because of 
its $\cos \delta$ dependence. 
Careful spectrum analysis would be required to resolve the confusion between 
the matter-CP and phase-CP effects.

\item 
For smaller $s_{13} \simeq \epsilon$, which may be relevant for neutrino factory 
setting, one can show by using the $\epsilon$ perturbation theory \cite{NSI-perturbation} 
with small matter density variation of $r^{A}_{n} \sim \epsilon$  that the terms 
sensitive to density variation are at most of order $\epsilon^3$.
It confirms qualitatively the conclusion reached in \cite{ota-sato01}.

\end{itemize}

\section*{Acknowledgments}

I thank Shoichi Uchinami for his useful comments.
This work was supported in part by KAKENHI, Grant-in-Aid for 
Scientific Research, No 19340062, Japan Society for the 
Promotion of Science.

\end{document}